\documentclass[11pt]{article} 

\usepackage{amscd,amsmath}
\usepackage {amsfonts} 
\usepackage{epsfig,graphicx,graphics}
\usepackage{citesort}
\usepackage{cancel}
\DeclareGraphicsExtensions{.pdf,.png,.jpg}

\title{ Equivalent Theories Redefine {Hamiltonian} Observables to Exhibit Change in General Relativity}

\author{J. Brian Pitts \\  
 University of Cambridge  \\  \\   \emph{Classical and Quantum Gravity} {\bf 34} (2017) 055008,\\ doi.org/10.1088/1361-6382/aa5ce8.} 

\date{30 January 2017}  

\begin{document}

\maketitle

\pagebreak

\abstract{Change and local spatial variation are missing in canonical General Relativity's observables as usually defined, an aspect of the problem of time.  Definitions can be tested using equivalent formulations of a theory, non-gauge and gauge, because they must have equivalent observables and everything is observable in the non-gauge formulation.  Taking an observable from the non-gauge formulation and finding the equivalent in the gauge formulation, one requires that the equivalent be an observable, thus constraining definitions.   For massive photons, the de Broglie-Proca non-gauge formulation observable $A_{\mu}$ is equivalent to the Stueckelberg-Utiyama gauge formulation quantity  $A_{\mu} + \partial_{\mu} \phi,$ which must therefore be an observable.  To achieve that result, observables must have $0$ Poisson bracket not with each first-class constraint, but with the Rosenfeld-Anderson-Bergmann-Castellani gauge generator $G$, a tuned sum of first-class constraints, in accord with the Pons-Salisbury-Sundermeyer definition of observables.

 The definition for external gauge symmetries can be tested using massive gravity, where one can install gauge freedom by parametrization with clock  fields $X^A.$ The non-gauge observable $g^{\mu\nu}$ has the gauge equivalent $ X^A,_{\mu} g^{\mu\nu} X^B,_{\nu}.$ The Poisson bracket of $ X^A,_{\mu} g^{\mu\nu} X^B,_{\nu}$ with $G$ turns out to be not $0$ but a Lie derivative. This non-zero Poisson bracket refines and systematizes Kucha\v{r}'s proposal to relax the $0$ Poisson bracket condition with the Hamiltonian constraint.   Thus observables need covariance, not invariance, in relation to external gauge symmetries. 

 The Lagrangian and Hamiltonian for massive gravity are those of General Relativity + $\Lambda$ + 4 scalars, so the same definition of observables applies to General Relativity.  Local fields such as $g_{\mu\nu}$ are observables.  Thus observables change. Requiring equivalent observables for equivalent theories also recovers Hamiltonian-Lagrangian equivalence. }


\section{Problems of Time and Space }

There has long been a problem of missing change in observables in the constrained Hamiltonian formulation of General Relativity (GR) \cite{BergmannGoldberg,AndersonChange,KucharCanadian92,IshamTime,KucharCanonical93}.
The typical definition is that  observables  have $0$ Poisson bracket with all first-class constraints  \cite{Bergmann,DiracLQM,Govaerts,HenneauxTeitelboim,RotheRothe}.
   This problem of missing change owes much to the condition  $\{O, \mathcal{H}_0\} =0$.  There is also a  problem of space:  local spatial variation is excluded by the condition for observables  $\{O, \mathcal{H}_i\} =0$, pointing to global spatial integrals instead   \cite{TorreObservable}.

The definition of observables is not uncontested.  Bergmann himself offered a variety of inequivalent definitions, essentially Hamiltonian or not, local or not \cite{BergmannKomarRoyaumont,Bergmann,BergmannHandbuch}. Here he envisaged locality:
\begin{quote} General relativity was conceived as a local theory, with locally well defined physical characteristics.  We shall call such quantities \emph{observables}. \ldots We shall call \emph{observables} physical quantities that are free from the ephemeral aspects of choice of coordinate system and contain information relating exclusively to the physical situation itself.  Any observation that we can make by means of physical instruments results in the determination of observables\ldots  \cite[p. 250]{BergmannHandbuch}. \end{quote} 
In pursuit of Hamiltonian-Lagrangian equivalence,  Pons, Salisbury and Sundermeyer have proposed a reformed definition of observables using the Rosenfeld-Anderson-Bergmann-Castellani gauge generator $G$, a tuned sum of all first-class constraints including the primaries \cite{AndersonBergmann,CastellaniGaugeGenerator,PonsSalisburySundermeyerFolklore}.

Kucha\v{r}, seeking real change, dropped the condition $\{ O, \mathcal{H}_0 \} =0$  \cite{KucharCanadian92,KucharCanonical93}.  Smolin has insisted that ``observables'' be really observable:  
\begin{quote} 
\ldots  we must ask if any of the observables are actually measurable
by observers who live inside the universe. If they are not then we cannot
use the theory to actually explain or predict any feature of our universe
that we may observe. If we cannot formulate a cosmological theory in
terms that allow us to confront the theory with things we observe we are
not doing science\ldots. 
 And the worrying fact is that none of the quantities which we have control over, as formal observables, are in fact
measurable by us. We certainly have no way to measure the total spacetime
volume of the universe or the spacetime average of some field.  \cite[p. 115]{SmolinPresent}  \end{quote}
According to Kiefer, 
\begin{quote} 
Functions $A(q,p)$ for which $\{A, \phi_A\} \approx 0$ holds are often called \emph{observables} because they do not change under a redundancy transformation.  It must be emphasized that there is no a priori relation of these observables to observables in an operational sense.  This notion was introduced by Bergmann in the hope that these quantities might play the role of the standard observables in quantum theory (Bergmann 1961). 
 \cite[p. 105; see also p. 143]{Kiefer3rd} \end{quote} 
The failure of observables to play their expected role has also led to circumvention with new concepts \cite{RovelliPartialObservables,DittrichPartialConstrained,TamborninoObservables}.


By requiring that  empirically equivalent theories have equivalent observables using the novel examples of  massive electromagnetism and  massive gravity, this paper shows the need for two reforms of the definition of observables:   $G$ rather than separate first-class constraints  \cite{PonsSalisburySundermeyerFolklore}  \emph{and} a novel non-zero Lie derivative Poisson bracket for external symmetries, partly inspired by Kucha\v{r}.  As a result, observables are local $4$-dimensional scalars, vectors, tensors, densities, \emph{etc}. (on-shell), as in Lagrangian/geometric formulations, including the metric tensor.  Thus change and local spatial variation are present after all.


\section{First-Class Constraints and Gauge?}

The original view about the precise relationship between  first-class constraints and gauge freedom retained manifest equivalence to the Lagrangian, as in Rosenfeld and Anderson and Bergmann \cite{RosenfeldQG,AndersonBergmann,SalisburySundermeyerRosenfeldQG}.  This view disappeared later in the 1950s in favor of distinctively Hamiltonian ideas with no Lagrangian equivalent---ideas that treated the constraints separately rather than as a team, or that dropped the primary constraints, or that extended the Hamiltonian with first-class secondary constraints.  The recovery of manifest Hamiltonian-Lagrangian equivalence started in  the later 1970s 
 \cite{MukundaSymmetries,MukundaGaugeGenerator,CastellaniGaugeGenerator,ShepleyPonsSalisburyTurkish}.  It holds that gauge transformations are generated by a 
 \emph{tuned sum} of first-class  constraints (primary, secondary, \emph{etc.}), the ``gauge generator'' $G$.  
For Maxwell's electromagnetism  \begin{eqnarray}  G(t)= \int d^3x (-\dot{\epsilon}(t,x) \pi^0 + \epsilon(t,x) \pi^i,_i(t,x)). \end{eqnarray}
  $ G$ involves two first-class constraints, but smeared with only one arbitrary function and its (negative) time derivative. $G$ performs as expected on  $A_{\mu}$:   $ \{ G(t), A_{\mu}(t,y) \} =  \epsilon,_{\mu}(t,y). $
This $G$-based view competes with what became the usual view  $ \{ FC, O\}, $ that each first-class constraint  $FC$ \emph{alone}  generates a gauge transformation. 

Fortunately one  can \emph{test} definitions  by calculation using two formulations of a theory, one without gauge freedom and one with gauge freedom.  The formulations, being empirically equivalent, must have equivalent observables, if observables deserve their name.  The equivalence of non-gauge and gauge formulations of massive quantum electrodynamics is presupposed in quantum field theory to show that the theory is renormalizable (shown using the Stueckelberg-Utiyama gauge formulation with a gauge compensation field) and unitary (shown using in effect the de Broglie-Proca formulation)  \cite[pp. 738, 739]{PeskinSchroeder}\cite[chapter 21]{WeinbergQFT2}\cite[chapter 10]{Kaku}.  

 
 \section{Massive Electromagnetisms and Equivalence}

Using non-gauge and gauge formulations of massive electromagnetism, one can test definitions by calculation.  
The de Broglie-Proca massive electromagnetic theory has a photon mass term $ -\frac{1}{2}m^2 A^{\mu} A_{\mu}$.   By contrast there is Maxwell-like gauge freedom in Stueckelberg-Utiyama massive electromagnetism  \cite{Ruegg,UtiyamaMesonGravity1}
   with the mass term  \begin{eqnarray} - \frac{m^2}{2} (A_\mu + \partial_\mu \phi) (A^\mu+ \partial^\mu \phi). \end{eqnarray} 
 $\phi$ is the gauge compensation ``Stueckelberg field,'' which allows the gauge transformation $A_{\mu} \rightarrow A_{\mu} + \partial_{\mu} \psi,$ $\phi \rightarrow \phi - \psi$ so that  $A_{\mu} +\partial_{\mu} \phi \rightarrow A_{\mu} +\partial_{\mu} \phi.$
Massive electromagnetism approaches massless (Maxwell) as $m \rightarrow 0,$ both classically and in quantum field theory   \cite{BelinfanteProca,Glauber,BoulwareYM,SlavnovFaddeev,GoldhaberNieto2009}.  

\vspace{.2in} 
 
\begin{center}
{\bf  Testing Definitions  with Massive Electromagnetisms }  
\end{center}
$$\begin{CD}
\text{{\bf de Broglie-Proca}}  @.    \text{{\bf Stueckelberg-Utiyama}} \\
 \mathcal{L}: A_{\mu} \text{ observable}    @>install>gauge>  \mathcal{L}: A_{\mu}+\partial_{\mu} \phi  \text{ observable} \\
@VconstrainedVLegendreV                                       @VconstrainedVLegendreV \\
\mathcal{H}:  A_{\mu}, \pi^i \text{ observable} @>demand>equivalence>  \mathcal{H}:  \{O,FC\}=0  \\ 
 \text{ because no FC constraints}      @.                \text{or } \{O,G\}=0? 
\end{CD}$$

\vspace{.15in} 


One can test definitions  by seeing which one satisfies the demand of equivalent observables. 
 Non-gauge de Broglie-Proca massive electromagnetism has \emph{no} first-class constraints, but only second-class constraints  \cite{Sundermeyer,GitmanTyutin}. The primary constraint is $\pi^0 = \frac {\partial \mathcal{L}_p }{\partial A_{0,0} } = 0.$ The constrained Hamiltonian is  \begin{eqnarray}  \mathcal{H}_{p} =  \frac{1}{2}(\pi^a)^2 + \pi^a A_{0,a} +\frac{1}{4} F_{ij}F_{ij} + \frac{m^2}{2} A_i^2  - \frac{m^2}{2} A_0^2. \end{eqnarray}    Preserving the primary constraint gives the secondary constraint, a modified phase space form of the Gauss law:  $\{ \pi^0(y),  \int d^3 x \mathcal{H}_{pc} (x) \} = \pi^a ,_a (y) + m^2 A_0 (y) = 0.$    Taking the Poisson brackets of the constraints among themselves, one gets  $\{ \pi^0 (x), \pi^a ,_a + m^2 A_0 (y) \} = -m^2 \delta(x,y), $  $ \{ \pi^0 (x), \pi^0 (y) \} =0,$ and $\{ \pi^a ,_a + m^2 A_0 (x), \pi^i ,_i + m^2 A_0 (y) \}=0.$  
Using either definition, everything is observable including  $A_i(x)$, $\pi^i(x)$ and $A_0(x)$ (though $A_0$ is redundant), because there are no first-class constraints. 


\subsection{Gauge Massive Electromagnetism}

The Hamiltonian treatment of Stueckelberg-Utiyama  gauge massive electromagnetism might be novel.  
One defines canonical momenta and finds the usual primary constraint  $\pi^0 = \frac {\partial \mathcal{L}_{s} }{\partial A_{0,0} } = 0$ and the usual momenta $\pi^i$ conjugate to $A_i$. 
But there is  a new momentum $P = \frac {\partial \mathcal{L}_{s} }{\partial \phi,_0 } = m^2 A_0 + m^2 \phi,_0$ for the gauge compensation field $\phi.$ One gets the Hamiltonian  \begin{eqnarray}  \mathcal{H}_{s} =  \frac{\pi^{a2} }{2} + \pi^a A_{0,a} +\frac{F_{ij}^2}{4}  + \frac{m^2 A_i^2}{2}  + \frac{P^2}{2m^2} -A_0 P +m^2 A_i \phi,_i + \frac{m^2 \phi,_i^2}{2}. \end{eqnarray}  Preserving the primary constraint $\pi^0 =0$ yields the secondary constraint $ \pi^a ,_a + P = 0.$    All constraints are first-class: $\{ \pi^0 (x), \pi^0(y) \} = 0,$  $\{  \pi^a,_a + P(x), \pi^b,_b + P(y) \} = 0,$  and crucially $\{ \pi^0 (x), \pi^a,_a + P(y) \} = 0,$  as in Maxwell's theory.

The gauge generator $G$ is  $ G= \int d^3x (-\dot{\epsilon} \pi^0 + \epsilon [\pi^i,_i + P]).$
$G$ changes the canonical action $\int dt \int d^3x [ p(x) \dot{q}(x) -\mathcal{H}]$ by at most a boundary term (also see \cite{FirstClassNotGaugeEM}), so Hamilton's equations are gauge-invariant under $G$, but not under the separate  first-class constraints.

Now one can ascertain whether observables $O$  need to satisfy both
 $\{O,   \int d^3x \xi(t,x) \pi^0(x)  \}=0$ and $\{ O,  \int d^3x \epsilon(t,x) [\pi^i,_i + P](x) \}=0$, or  merely $\{O, G\} =0$.  The Stueckelberg-Utiyama expression $A_{\mu} + \partial_{\mu} \phi$ is equivalent to the de Broglie-Proca field $A_{\mu}.$   The latter is an observable. Hence  $A_{\mu} + \partial_{\mu} \phi$ must be an observable.   The primary constraint acting on  $A_{\mu} + \partial_{\mu}\phi$ gives 
\begin{eqnarray}  \{ \int d^3x \xi(t,x) \pi^0(x), A_{\mu}(y) + \partial_{\mu} \phi (y) \}  = -\xi(y) \delta^0_{\mu}\neq 0 \end{eqnarray}  for $\mu=0$.
The secondary gives \begin{eqnarray}  \{ \int d^3x \epsilon(t,x) [\pi^k,_k(x) + P(x)], A_{\mu}(y) + \partial_{\mu}\phi (y) = - \dot{\epsilon} \delta^0_{\mu} \end{eqnarray}  after cancellation and using the Anderson-Bergmann velocity Poisson bracket   $ \{\dot{q}, F\} = \frac{\partial}{\partial t} \{q,F \}$ \cite{AndersonBergmann}.  Happily,  $A_{i} + \partial_{i} \phi$ is observable on either definition. 
Strikingly, $ A_0 + \dot{\phi}$ is not observable for separate first-class constraints   definition, but it \emph{is} observable using the $G$ definition  \begin{eqnarray} \{ G, A_{\mu} + \partial_{\mu} \phi \}= \dot{\epsilon} \delta_{\mu}^0 - \dot{\epsilon} \delta_{\mu}^0=0, \end{eqnarray}    
thus providing equivalent observables for equivalent theories.

One might think that $ A_0 + \dot{\phi}$ does not need to be observable, because it is equal to $ m^{-2} P$  using the Hamilton equation $ \dot{\phi} = \frac{ \delta H_s}{\delta P} = m^{-2} P -A_0.$  However, that equation itself is not preserved by each first-class constraint, whereas it is preserved by the gauge generator $G$.  
One has for the primary constraint 
\begin{eqnarray} 
\{     \int d^3x \xi(t,x) \pi^0(x)      ,  \dot{\phi} - \frac{ \delta H_s}{\delta P}(y) \}   =  - \xi(t,y) \neq 0,   
\end{eqnarray} 
proving the point. 
Thus one cannot appeal to $\dot{\phi} - \frac{ \delta H_s}{\delta P}=0$ without rejecting the view that the primary first-class constraint by itself generates a gauge transformation, the foundation of the typical definition of observables. 
 One can also consider the secondary constraint, which requires the Anderson-Bergmann velocity Poisson bracket \cite{AndersonBergmann} $\{ \frac{\partial q}{\partial t}, F\} = \frac{\partial}{\partial t} \{ q, F\}$ (which takes precedence over the product rule for Poisson brackets): 
\begin{eqnarray} 
\{ \int d^3x \epsilon(t,x) [\pi^i,_i + P](x),  \dot{\phi} - \frac{ \delta H_s}{\delta P}(y) \}  = -\dot{\epsilon}(y).
\end{eqnarray} 
This term from the secondary constraint cancels that from the primary constraint in $G$:
\begin{eqnarray} 
\{ G,  \dot{\phi} - \frac{ \delta H_s}{\delta P}(y) \} =  \dot{\epsilon}(y) -\dot{\epsilon}(y) =0. 
\end{eqnarray} 

The Anderson-Bergmann velocity product rule  $ \{\dot{q}, F\} = \frac{\partial}{\partial t} \{q,F \}$ played an important role in ensuring that equivalent formulations have equivalent observables.  That role provides support for the correctness and importance of this bracket, which is usually neglected. This rule has the surprising consequence that one version of the Poisson bracket product rule---the case when the velocity is isolated---fails in order to respect the time differentiation product rule:
\begin{eqnarray}
 \{  \dot{q}, FG \} \neq  \{ \dot{q}, F \} G + F \{ \dot{q}, G \},  \hspace{.08in} {because} \nonumber \\
\{ \dot{q}, FG \} = \frac{ \partial \{ q, FG \}  }{ \partial t} \; by \;  Anderson-Bergmann \; velocity \; bracket \nonumber \\ =   \frac{ \partial }{ \partial t}  ( F\{ q, G \} + \{q, F\} G  )   \; by \; Poisson \; product \; rule  \nonumber \\
 = \frac{ \partial   F  }{ \partial t} \{ q, G \} + F \frac{ \partial    \{ q, G \}  }{ \partial t} 
+    \{ q, F \} \frac{ \partial   G  }{ \partial t} +  \frac{ \partial    \{ q, F \}  }{ \partial t} G \; by \; derivative \; product \; rule \nonumber \\  %
= \frac{ \partial   F  }{ \partial t} \{ q, G \} + F    \{ \dot{q}, G \}   
+    \{ q, F \} \frac{ \partial   G  }{ \partial t} + \{ \dot{q}, F \} G \; by \; Anderson-Bergmann. 
\end{eqnarray} 
It appears that a fundamental theory of these matters is still lacking. 



\section{Internal \emph{vs.} External Gauge Symmetries and Invariance \emph{vs.} Covariance }

There is no guarantee that external gauge symmetries behave in the same way as internal symmetries, so possibly the definition  $\{ O, G \}=0$ needs modification for external symmetries.  In  GR,  $G$ acting on $g_{\mu\nu}$  gives the $4$-d Lie derivative  \cite{CastellaniGaugeGenerator},   $ \pounds_{\xi} g_{\mu\nu} = \xi^{\alpha} g_{\mu\nu},_{\alpha} + g_{\mu\alpha} \xi^{\alpha},_{\nu} + g_{\alpha\nu} \xi^{\alpha},_{\mu}.$ The second and third terms are (for weak fields in nearly Cartesian coordinates) analogous to the electromagnetic case, but the transport term $ \xi^{\alpha} g_{\mu\nu},_{\alpha}$ differentiates $g_{\mu\nu}$, thus making the transformation  ``external''   \cite{BergmannNonlinear}.

The definition $ \{ O, G \} =0$  does not yield local observables in GR.  Because $G$ gives the $4$-dimensional Lie derivative, the definition of observables $\{ O,  G[\xi^{\alpha}] \}=0$ $(\forall \xi^{\alpha})$ implies $\pounds_{\xi} O =0$  $(\forall \xi^{\alpha})$: the directional derivative of $O$ vanishes along any vector field. 
  Can one devise a systematic definition of observables with local change, and that does not depend essentially on the Hamiltonian formalism?

One might amend Kucha\v{r}'s proposal in two ways (apart from using $G$ \cite{PonsSalisburySundermeyerFolklore}).  First, Kucha\v{r}'s common-sense argument against $\{ O, \mathcal{H}_0 \} =0$ is equally persuasive  against $\{ O, \mathcal{H}_i \} =0$ (which he retains).   Second, abolishing all  restrictions on time gauge behavior is  unnecessarily strong.  There is an  intermediate position, not invariance but covariance, imposing some coordinate transformation rule (scalar, vector, \emph{etc.}).  Infinitesimally, one would thus expect (especially after embracing $G$) a $4$-dimensional Lie derivative, not $0$, as the result of the Poisson bracket.  Whereas electromagnetic gauge transformations are ineffable mental acts with no operational correlate (no knob or reading on a voltmeter), necessitating invariance, general relativistic gauge (coordinate) transformations are already familiar from geography and daylight savings time  and only require a transformation rule (covariance).  



\section{Testing Definitions using Massive Gravity} 
 
Massive gravity can have gauge freedom (re)installed by  parametrization, promoting  preferred coordinates into fields varied in the action principle  
\cite{Kuchar73,Schmelzer,Arkani,MassiveGravity1}.  
 Massive gravity was found in the early 1970s to imply  either instability (spin $2$-spin $0$ ``ghost'') or a discontinuous massless limit \cite{vDVmass1,Zakharov,vDVmass2,DeserMass} or both.  Progress was made on both fronts during the 2000s \cite{Vainshtein2,deRhamGabadadze,HassanRosenNonlinear}, along with new challenges \cite{DeserWaldronAcausality}.  Fortunately, for present purposes it doesn't matter  what problems massive gravity has---even ``ghost'' theories are acceptable, because physical viability and certainly quantization are not in view.  What matters is the relationship between the non-gauge and gauge versions and their following from  variational principles.  Because physical reasonableness is irrelevant, one might as well choose the version that makes the calculations the easiest.  The Freund-Maheshwari-Schonberg (FMS) theory  \cite{FMS}, when parametrized, gives \emph{minimally} coupled  scalar clock fields, as Schmelzer noted \cite{Schmelzer}.

The observables in  non-gauge massive gravity are obvious because all constraints are second-class \cite{PittsQG05}. Everything is observable, including the $4$-metric $g_{\mu\nu}$ and the  non-zero momenta $\pi^{mn}.$ Most non-gauge massive gravities, including this one,  are  merely Poincar\'{e}-invariant \cite{FMS,OP,HassanRosenNonlinear}. 
The FMS mass term is \begin{eqnarray} \mathcal{L}_{m} = m^2  \sqrt{-g} + m^2 \sqrt{-\eta} - \frac{1}{2} m^2 \sqrt{-g} g^{\mu\nu} \eta_{\mu\nu}, \end{eqnarray}  where  $\eta_{\mu\nu}=diag(-1,1,1,1)$ in  Cartesian coordinates.  %

One obtains the gauge version by parametrization of the  mass term.  Now the reason for choosing the FMS theory becomes evident: its mass term, in contrast to the others  (\emph{e.g.}, \cite{OP,HassanRosenNonlinear}), gives \emph{minimally coupled} scalar clock fields in \begin{eqnarray}  \sqrt{-g} g^{\mu\nu} \eta_{AB} \frac{\partial X^A}{\partial x^{\mu}}  \frac{\partial X^B}{\partial x^{\nu}}, \end{eqnarray} instead of using inverses, determinants, and/or fractional powers of $\eta_{AB} \frac{\partial X^A}{\partial x^{\mu}}  \frac{\partial X^B}{\partial x^{\nu}},$ or sums thereof. 
(Surprisingly, one can actually do calculations in the general case \cite{KlusonHamiltonian}.) 
The parametrized mass term is \begin{eqnarray} \mathcal{L}_{mg} = m^2  \sqrt{-g} + m^2 \sqrt{-\eta} - \frac{1}{2} m^2 \sqrt{-g} g^{\mu\nu} \eta_{AB} \frac{\partial X^A}{\partial x^{\mu}}  \frac{\partial X^B}{\partial x^{\nu}}, \end{eqnarray} where $\eta_{AB}=diag(-1,1,1,1).$ 



Knowing that  $g^{\mu\nu}$ is observable in the non-gauge theory, one should  \emph{require} that observables in gauge massive gravity  be equivalent to the non-gauge observables. The equivalent quantity is $g^{\mu\nu} X^A,_{\mu} X^B,_{\nu} :$ the clock field gradients act as the tensor transformation law to the preferred Cartesian coordinates.  By seeing how  $g^{\mu\nu} X^A,_{\mu} X^B,_{\nu} $ behaves, one learns how observables behave under \emph{external} gauge transformations.  


 \begin{center}  {\bf \hspace{.1in} Testing Definitions of Observables with Massive Gravities \vspace{-.2in}}  \vspace{-.2in}  \end{center}

$$\begin{CD}
\text{{\bf  Massive Gravity}}  @.    \text{{\bf  Parametrized Massive Gravity}} \\
 \mathcal{L}: g^{\alpha\beta} \text{ observable}    @>install>gauge>  \mathcal{L}: g^{\mu\nu} X^A,_{\mu} X^B,_{\nu}  \text{ observable} \\
@VconstrainedVLegendreV                                       @VconstrainedVLegendreV \\
\mathcal{H}:  g^{\alpha\beta},  \text{ momenta }  @.   \mathcal{H}: \cancel{  { \{O,FC\}=0 } } \vspace{-.1in}  \\   
 \text{observable because }     @>demand>equivalence>                \text{or } \{O,G\}=0 \vspace{-.1in} \\
\text{no FC constraints }    @.                                   \text{or } \{O, G \} \sim \pounds_{\xi} O? \vspace{-.1in}
\end{CD}$$


\subsection{Hamiltonian for Gauge Massive Gravity}  

For GR with  minimally coupled scalar fields  and $\Lambda$, the Poisson bracket `algebra' of constraints is just  as in GR \cite{Sundermeyer}.  For  parametrized  Freund-Maheshwari-Schonberg theory, the same result therefore holds.  It is straightforward to take the parametrized Lagrangian density and perform the constrained  Legendre transformation with 4 minimally coupled scalars and $\Lambda$   GR \cite{MTW,Sundermeyer,Wald}.  In the ADM $3+1$ split, the $4$-metric is broken into the  lapse $N$, the shift vector $\beta^{i},$ and spatial metric $h_{ij}.$  There are new canonical momenta for the clock fields:  $ \pi_A = \frac{\partial \mathcal{L}_{mg} }{\partial X^A,_0 }.$  Inverting, one gets  $\dot{X}^A = N  \pi_C \eta^{AC} m^{-2}/ \sqrt{h}  + \beta^i X^A,_i.$ 
  The Hamiltonian density is 
{ \begin{eqnarray} \mathcal{H}_{mg} = N  \left(\mathcal{H}_0    -m^2 \sqrt{h} + \frac{ \pi_A \pi_B \eta^{AB} }{2 m^2 \sqrt{h} } +  \frac{m^2}{2} \sqrt{h} h^{ij} X^A,_i X^B,_j \eta_{AB} \right) \nonumber \\ + \beta^i (\mathcal{H}_i+  X^A,_i \pi_A) - m^2 \sqrt{-\eta}, \end{eqnarray}  with   $\mathcal{H}_0$ and $\mathcal{H}_i$ as in GR.  }  This expression has the same form as in GR (apart from an irrelevant constant $\sqrt{-\eta}$) if one defines a total (gravitational plus matter) Hamiltonian constraint $\mathcal{H}_{0T}$ and a total momentum constraint $\mathcal{H}_{iT}$:  $ \mathcal{H}_{mg} = N \mathcal{H}_{0T}  + \beta^i \mathcal{H}_{iT} -m^2 \sqrt{-\eta}.$

  Avoiding velocities  requires $3+1$ split of coordinate transformation descriptor $\xi^{\mu}$  \cite{CastellaniGaugeGenerator,PonsSalisburyShepleyYang}:   $\epsilon^{\perp} = N \xi^0$ is primitive and so has $0$ Poisson brackets; the same holds for  $\epsilon^i = \xi^i + \beta^i \xi^0.$  The primary constraints are as in GR:  $p$  conjugate to $N$ and $p_i$  conjugate to $\beta^i$ both vanish.   
The generator of changes of time coordinate in GR is  \cite{CastellaniGaugeGenerator,PonsSalisburyShepleyYang} \begin{eqnarray} G[\epsilon, \dot{\epsilon}] = \int d^3x [\epsilon^{\perp} \mathcal{H}_0  + \epsilon^{\perp} p_j h^{ij} N,_i + \epsilon^{\perp}(N p_i h^{ij}),_j + \epsilon^{\perp} (p N^j),_j + \dot{\epsilon}^{\perp} p]. \end{eqnarray} 
This entity generates on phase space $\times$ time a transformation that, for solutions of the Hamiltonian field equations, changes the time coordinate in accord with  $4$-dimensional tensors.  Given how the gauge generator can be built algorithmically starting with the primary constraints \cite{PonsDirac}, one would expect the same expression for the gauge generator for parametrized massive FMS gravity, only with matter included in the secondary constraints.  The Hamiltonian takes the form of  GR + $\Lambda$ + minimally coupled scalars with altered matter-containing constraints $\mathcal{H}_{0T}$ and $\mathcal{H}_{iT}.$  One can verify that the resulting modified expression for $G$ indeed generates gauge transformations.  For the space-time metric there is no difference because matter does not couple to gravitational momenta.  For the new matter fields one has on-shell \begin{eqnarray}  \{ G[\epsilon, \dot{\epsilon}], X^A(y) \} = - \epsilon^{\perp}(y) \pi_B \eta^{BA} /(m^2 \sqrt{h}) =   - \xi^0 X^A,_0 + \xi^0 \beta^i X^A,_i.  \end{eqnarray}

  
 The spatial gauge generator for GR is \cite{CastellaniGaugeGenerator,PonsSalisburyShepleyYang} \begin{eqnarray}  G[\epsilon^i, \dot{\epsilon}^i] = \int d^3x [\epsilon^i \mathcal{H}_i  +  \epsilon^i N^j,_i p_j - \epsilon^j,_i N^i p_j  + \epsilon^i N,_i p + \epsilon^i,_0 p_i]. \end{eqnarray}  It generates $3$-d spatial  Lie derivatives of the $4$-metric $g_{\mu\nu}$ even off-shell.  Making the obvious alteration of the secondary constraint to include matter through  $\mathcal{H}_{iT}$  gives the correct gauge generator, giving a Lie derivative of the scalar clock fields:  \begin{eqnarray} \{ G[\epsilon^i, \dot{\epsilon}^i], X^A(y) \} = - (\xi^i + \beta^i \xi^0) X^A,_i. \end{eqnarray} 

The full gauge generator $G= G[\epsilon, \dot{\epsilon}] + G[\epsilon^i, \dot{\epsilon}^i]$ is the sum of these two parts \cite{CastellaniGaugeGenerator}.  Acting on the clock fields $X^A,$ it gives (on-shell) 
\begin{eqnarray} 
 \{ G[\epsilon, \dot{\epsilon}] + G[\epsilon^i, \dot{\epsilon}^i], X^A(y) \} = 
- \xi^0 X^A,_0  - \xi^i X^A,_i = - \xi^{\mu} X^A,_{\mu}: \end{eqnarray} 
the  $4$-dimensional expression for the Lie derivative of a scalar field with respect to the space-time vector field $\xi^{\mu}$ describing the infinitesimal coordinate transformation.  
 One  knows that  $ \{ G[\epsilon, \dot{\epsilon}] + G[\epsilon^i, \dot{\epsilon}^i], g^{\mu\nu} \} = - \pounds_{\xi} g^{\mu\nu} $ \cite{CastellaniGaugeGenerator} in GR.  It still holds here because the new material part of the total momentum constraint has no gravitational momenta. 

One can apply the full $G$ to  $g^{\mu\nu} X^A,_{\mu} X^B,_{\nu}$, the equivalent of the non-gauge observable $g^{\alpha\beta}$.  The calculation of $\{G, g^{\mu\nu} X^A,_{\mu} X^B,_{\nu}\}$  is effortless using  the Leibniz rule for Poisson brackets, the Anderson-Bergmann velocity Poisson bracket (for $0$ values of indices $\mu$ and $\nu$), some of Hamilton's equations (making the result valid only on-shell), the Leibniz rule for Lie derivatives, and the commutativity of Lie and partial derivatives \cite{Yano}. 
The result is  \begin{eqnarray} \{G, g^{\mu\nu} X^A,_{\mu} X^B,_{\nu}\} = -\pounds_{\xi} (g^{\mu\nu} X^A,_{\mu} X^B,_{\nu}). \end{eqnarray} By the equivalence of the non-gauge and gauge observables, $g^{\mu\nu} X^A,_{\mu} X^B,_{\nu}$ must be observable in the gauge theory because $g^{\mu\nu}$ is observable in the non-gauge theory.  Thus observables  give a Lie derivative rather than $0$ under  Poisson bracket with $G$. The widely received $0$ Poisson bracket condition is incorrect in at least this application.

As it happens, $g^{\mu\nu} X^A,_{\mu} X^B,_{\nu}$ is a scalar under change of arbitrary coordinates $x^{\mu}.$ The Lie derivative is just a directional derivative in this case; the full notion of a Lie derivative is used only in intermediate stages of the calculation when $g^{\mu\nu} X^A,_{\mu} X^B,_{\nu}$is broken into its factors.   The technology of Lie derivatives, however, works perfectly well for pseudo-scalars, vectors, tensors, densities, connections, \emph{etc.}  Hence there seems to be no difficulty in requiring mere covariance, not invariance, under an external gauge symmetry. So there is no obvious objection to regarding pseudoscalars, vectors, tensors, densities, connections, \emph{etc}. as observables once one admits scalars.  The numerical components, of course, will be relativized to a coordinate system in the familiar rule-governed way.

\section{Are Observables Too Easily Had?}

One might worry, however, that Lie derivatives, and hence on the above definition observables, are too easily had. Because Lie differentiation commutes with partial differentiation, such non-tensorial entities as $g^{\mu\nu},_{\alpha}$ have a Lie derivative. Indeed one has 
\begin{eqnarray} \{ G[\epsilon, \dot{\epsilon}] + G[\epsilon^i, \dot{\epsilon}^i], g^{\mu\nu},_{\alpha} \} = - \pounds_{\xi} g^{\mu\nu},_{\alpha}. \end{eqnarray}
Is the   coordinate-dependent `quantity' $g^{\mu\nu},_{\alpha}$, which has no transformation rule relating its manifestations in different coordinate systems, to be regarded as an observable?   To prevent such an absurd conclusion, one should further specify that the resulting Lie derivative be the kind of Lie derivative that  geometric objects have.  That involves a group property \cite{BergmannNonlinear}, which $ \pounds_{\xi} g^{\mu\nu},_{\alpha}$ lacks but $\pounds_{\xi} g^{\mu\nu}$ has.    Here a geometric object means a set of components in every admissible coordinate system and a transformation rule that relates the sets of components \cite{Nijenhuis,Schouten,Trautman}.  It isn't necessary that $ \pounds_{\xi} O$ itself be a geometric object, because a nonlinear geometric object has the property that its Lie derivative is not a geometric object, but the Lie derivative and the object itself together do form a geometric object  \cite{SzybiakLie,LaptevLie}.\footnote{It is slightly tricky that some interesting nonlinear would-be geometric objects, such as the symmetric square root of an indefinite metric tensor, actually fail to be defined for some coordinate systems  \cite{PittsSpinor,TAM2013TimeandFermionsConfProc,DeffayetSymmetricTetrad}. Such an entity is useful both for massive gravity theories and for spinors.  Hence the ``admissible'' coordinate systems might be restricted in such cases. }  If $ \pounds_{\xi} O$ must be the kind of Lie derivative that a geometric object has, then presumably $O$ itself is a geometric object, or at least acts like one for small coordinate transformations.  Thus observables turn out to be on-shell just $4$-dimensional differential geometry all over again.

 
\section{Application to General Relativity}

There is a strong case  that  the same definition of observables should apply in canonical GR, or, minimally, that this definition is well motivated, extensionally correct (giving the right instances), and simple.  First, the calculation with massive gravity proves that the $0$ Poisson bracket condition does not apply in every case. With the widely received $0$ Poisson bracket now known to fail in one case involving an external gauge symmetry, it becomes  a serious open question whether the $0$ Poisson bracket condition applies to GR.  Kucha\v{r} has previously doubted the $0$ Poisson bracket regarding time-related gauge transformations in GR.  For parametrized massive gravity, that condition is  disproved for temporal and spatial transformations as well.

It is not easy to point to any crucial difference between parametrized massive gravity and GR that should lead to different definitions of observables.  Parametrized massive gravity and GR with four minimally coupled scalar fields (one of them wrong-sign) and a cosmological constant have the \emph{same Euler-Lagrange equations and the same Lagrangians and Hamiltonians}, at least up to terms that do not affect the field equations.  If the Euler-Lagrange equations are not sufficient to specify the definition of observables, then a very precise and principled distinction is needed.  It seems doubtful that such a distinction exists. Unless such a distinction is proposed, the same definition is appropriate for GR.

Furthermore, this $4$-dimensional Lie derivative condition is a very natural condition from the standpoint of $4$-dimensional differential geometry.  The gauge transformations of General Relativity are just $4$-dimensional coordinate transformations. (As is traditional in constrained Hamiltonian GR \cite{BergmannNonlinear}, all transformations are construed passively.)  A Poisson bracket with the gauge generator $G$ should generate infinitesimal coordinate transformations.  But infinitesimal coordinate transformations are, on technical grounds, presented with an additional transport term to arrive at the Lie derivative  \cite{BergmannLectures}.   Thus one would expect a Poisson bracket with $G$ to give a $4$-dimensional Lie derivative.  So it does, given the definition of observables above.  Thus the definition does what  one would expect  a definition to do, being well motivated and giving the expected examples.  

It is worthwhile to recall how the classical definition of the Lie derivative arises, as described by Bergmann.  \begin{quote}  Because the transformed $y_A$ are not compared with the original values at the same world point, but with the original values at that world point which possesses the same coordinate values prior to the transformation, [there arises]  a ``transport'' term  [$-y_A,_{\mu} \xi^{\mu}$].  \cite{BergmannNonlinear} \end{quote} 
This is the infinitesimal analog of comparing 1 a.m. Daylight Savings Time with 1 a.m. Standard Time an hour later: one compares different space-time points with the same numerical coordinate values in different coordinate systems.   Commuting with Lie differentiation is the valuable property that justifies using  such a physically peculiar ``fixed coordinate variation''  \cite{BergmannLectures}.  This issue was understood already as early as the 1910s \cite{WeylAction,KleinGREnergy1918,Noether} \cite[p. 271]{Landau}.  Weyl was very clear in 1917:  to get the transport term 
\begin{quote} I take the difference of $g^{ik}$ and $\bar{g}^{ik}$ at two space-time points, the second of which has the same co\"{o}rdinate values in the new co\"{o}rdinate system as the first in the old one; in other words, I perform a virtual displacement. \cite{WeylAction} 
\end{quote}
 Bergmann's statement (like Weyl's) is correct about the passive viewpoint, which is coherent and economical, so active interpretations do not require consideration.  The active viewpoint makes use of a dragging of field values, replacing the comparison of actual field values at different space-time points with a comparison of real and fictitious values at the same point, a lateral rather than progressive move.

The $0$ Poisson bracket condition, by implying vanishing Lie derivative in every direction, simply \emph{feeds in} a requirement of constancy for observables.  It is therefore little wonder that constancy has reappeared as a result.  But (the infinitesimal analog of) being the same at 1 a.m. Daylight Savings Time and at 1 a.m. Standard Time (an hour later) simply has nothing  to do with being free of gauge dependence or with being observable.  One can see this, for example, with  tea kettles:  if a tea kettle is boiling at 1 a.m. Daylight Savings Time but not boiling at 1 a.m. Standard Time, the tea kettle does not become unobservable in any sense, nor even difficult to observe; its boiling or not is not gauge-dependent, because different world-points are in view.   Being \emph{exactly} the same at different times might even exclude being observable:  physically real properties (besides physical constants), even such stable ones as the shape of a building, change a little bit over time because real materials are not infinitely rigid.

The indefinite signature of the background metric in massive gravity  plays no role in the argument.  What matters is that the background structure is strong enough to remove the gauge freedom.  That would also happen if the background structure were positive definite.  Then the parametrized theory would be like GR + $\Lambda$ +  four positive-energy minimally coupled scalar fields.   Again the Lie derivative result would hold.  Adding  $\Lambda$ and  minimally coupled scalars to GR does not change its Hamiltonian properties  \cite{Sundermeyer}. So the definition of observables should not change when the scalars and/or $\Lambda$ is  removed.

These considerations establish a strong presumption that the same definition applies to GR.  It isn't quite a proof.  But then no definition of observables in GR has been proven before, and likely none could be unless appeals to common sense (which shows change and local spatial variation) and Hamiltonian-Lagrangian equivalence are made, and to simplicity as well.  Those considerations also call for the Lie derivative with $G$ as the definition of observables \cite{GRChangeNoKilling}.  In a $4$-dimensional Lagrangian context, no one worries that scalar fields, vector fields, tensor fields, \emph{etc.} might be unobservable or gauge-dependent.  There is no distinctively Hamiltonian gauge freedom, if one succeeds in preserving Hamilton's equations including
$\dot{q} = \frac{\partial H}{\partial p}$ \cite{CastellaniGaugeGenerator,PonsDirac,FirstClassNotGaugeEM}.  Neither should there be any distinctively Hamiltonian problem of observables unless one makes distinctively Hamiltonian definitions with no good Lagrangian equivalent. There are only the old problems already handled by $4$-dimensional tensor calculus, now with some auxiliary fields (the nonzero canonical momenta) $\pi^{ab}$ in the canonical Lagrangian $p \dot{q}-H.$

In short, there is no reason to deny  that in GR, the $4$-metric  at a \emph{physically individuated} point is an observable, as are such concomitants as the curvature tensors---even in the Hamiltonian formulation (on-shell).  Individuating points is of course non-trivial, typically requiring four scalar fields, whether built from curvature or built from matter; thus it can take five scalar fields to observe one scalar field \cite{BergmannHandbuch,RovelliObservable}.  But that is not a distinctively Hamiltonian problem.  Neither does it  get in the way of $4$-dimensional tensor calculus:  the hoary rule
\begin{eqnarray}
g_{\rho^{\prime}\sigma^{\prime}}(p) = g_{\mu\nu}(p)  \frac{ \partial x^{\mu} }{\partial x^{\rho^{\prime} } }(p)  \frac{ \partial x^{\nu} }{\partial x^{\sigma^{\prime} } } (p)
\end{eqnarray} applies at any and every point $p$, notwithstanding any labors involved in finding a specific point.   Thus change and local spatial variation are not missing, but appear in the Hamiltonian formalism (on-shell) exactly where they appear in the Lagrangian/$4$-dimensional geometric formalism.  Change and local spatial variation appear for observables appear once one defines observables using $G$ and the Lie derivative, because observables on-shell are just geometric objects (at least infinitesimally). 
Change is missing only when and where there is a time-like Killing vector field \cite{GRChangeNoKilling}.


\section{Relation to Definition by Pons, Salisbury and Sundermeyer}

The definition of observables given here agrees with that given by Pons, Salisbury and Sundermeyer for internal symmetries, while differing regarding external symmetries such as one finds in General Relativity.  It seems possible that the difference is not one of fact, as though one definition would be true and another false, but rather of convention, with different theoretical choices made.  Whereas my definition follows the old tradition of passive coordinate transformations, Pons, Salisbury and Sundermeyer employ active transformations in their construction of observables \cite{PonsSalisburySundermeyer}.  
It seems to me that my definition is clearer and more direct, because their definition has, roughly speaking, three copies of the gauge freedom.  If active diffeomorphisms are present (as in their formalism but not mine), it is fitting that observables be invariant under them.  But ultimately they find active diffeomorphisms to be a detour:  
\begin{quote}
We have shown that it is possible to construct, albeit in a formal way, observables in general relativity by employing a gauge fixing using a scalar coordinatization. In this way we obtain a new understanding as to why finding observables in generally covariant theories is such a difficult mathematical task. But once these two points --- the existence and the difficulty of construction of observables --- have been made, a new vision emerges: that constructing these observables through the use of active diffeomorphism-induced symmetry transformations --- which are valid for every observer with his/her own coordinatization --- is not the most efficient procedure. Indeed, once we have proven that observables can be built for any observer, we can gladly dispose of this construction and just take the passive view of diffeomorphism invariance. We simply instruct each observer, having constructed his or her phase space solutions, to transform them to the intrinsic coordinate system!  We have indeed proven that the final result is coincident with the active construction.\ldots \\
Thus here is the guiding principle: \emph{let everyone adopt the same instrinsic coordinates}. Once this instruction is implemented all geometric onjects becomes observable! All observers attain the same description regardless of the coordinate system with which they begin their construction. In other words, the final description is invariant under alternations in this iniital arbitrary coordinate choice. \cite{PonsSalisburySundermeyerFolklore} 
\end{quote}  
Whereas they introduce active diffeomorphisms, achieve invariance under them, and then recognize their dispensability, in my approach active diffeomorphisms just never arise.  If three copies of the gauge freedom are present, it is quite appropriate if observables are invariant under two of them. 
But the choice of intrinsic coordinates is also arbitrary.  When one coordinate system is intrinsic, all are intrinsic---they simply use different functions of the Weyl scalars.  Because the world does not give us a unique choice for building coordinates out of the Weyl scalars, one would want to know how Pons-Salisbury-Sundermeyer observables transform under the choice of intrinsic coordinates. (By analogy, having everyone use Latin does not really achieve a convention-independent expression of thought.)  
 Presumably one would find only covariance under change of intrinsic coordinates.  My approach avoids both active diffeomorphisms and intrinsic coordinates, arriving directly at covariance (a transformation rule) under the one arbitrary gauge conventional choice, the space-time coordinates.  This approach might be extensionally equivalent to the Pons-Salisbury-Sundermeyer definition, arriving at geometric objects (or things that behave like them near the identity) as observables, and thus $4$-dimensional tensor calculus all over again.  The remarks on Einstein-Maxwell theory below are relevant, because both external and internal symmetries are involved.


\section{Conclusions} 
 
Finding that local tensor fields are observable satisfies Bergmann's and others' expectations  for locally varying observables.  Let us recall one of Bergmann's definitions. 
\begin{quote} General relativity was conceived as a local theory, with locally well defined physical characteristics.  We shall call such quantities \emph{observables}. \cite[p. 250, emphasis in the original]{BergmannHandbuch}.  \end{quote} 

One might wonder why it was necessary to lower standards for external symmetries to have a nonzero Poisson bracket. The Lie derivative has two kinds of terms with different origins.  If one could work with the non-numerical tensor-in-itself ${\bf g} = g_{\mu\nu} {\bf d}x^{\mu} \otimes {\bf d}x^{\nu}$, which is invariant, rather than the components  $g_{\mu\nu},$ which are covariant (in the sense of having a transformation rule), 
 one would  avoid the tensor transformation rule-induced correction terms   $  g_{\mu\alpha} \xi^{\alpha},_{\nu} + g_{\alpha\nu} \xi^{\alpha},_{\mu}$  present in the Lie derivative of the metric components.  If one could also avoid comparing different places, then one could avoid the analog of the transport term     $  \xi^{\alpha} g_{\mu\nu},_{\alpha} $ as well. Then in place of the Lie derivative there would be $0.$  Such a significant reworking of the component-based standard formalism would be an interesting but nontrivial project.  Until then, the component formalism with covariance and Lie derivatives will suffice.  The Hamiltonian formalism is manifestly spatially covariant, making the step from covariance to invariance perhaps not too large.  However, 
 the crucial role played by a time coordinate and the fact that tensorial time coordinate transformation behavior holds only on-shell \cite{FradkinVilkoviskyHLEquivalence,CastellaniGaugeGenerator,GRChangeNoKilling}, in contrast to manifest spatial covariance, suggest that achieving invariance rather than covariance for time and hence space-time will not be trivial in a traditional Hamiltonian formalism.

 Combining internal and external symmetries as in the Einstein-Maxwell theory  \cite{PonsSalisburyShepleyYang}, the arguments above imply that observables  are invariant under the internal gauge symmetry and covariant under the external symmetry, making $F_{\mu\nu}$ observable (though not for Yang-Mills) and $g_{\mu\nu}$   as well.   If one expected observables to have $0$ first-class constraint with the whole $G$, then the electromagnetic field would not be observable in Einstein-Maxwell theory because the external coordinate transformation part of $G$ changes $F_{\mu\nu}$ by a Lie derivative.  

A case not covered is supergravity, which mixes internal and external symmetries in a non-trivial way \cite{vanNReports}. 
 Finally, one should  explore the relationship between these reformed observables  and partial observables \cite{RovelliPartialObservables,DittrichPartialConstrained}.


\section{Acknowledgments}

This work was supported by John Templeton Foundation grants \#38761 and \#60745.




\end{document}